\documentclass{aa}
\usepackage{graphicx}
\begin{document}

\title{Rotating Skyrmion Stars}

\author{Rachid OUYED}

\institute{Nordic Institute for Theoretical Physics, Blegdamsvej 17,
DK-2100 Copenhagen, Denmark}
\offprints{ouyed@nordita.dk}

\date{Received/Accepted}

\abstract{
In a previous paper, using an equation of state  of
dense matter representing a fluid of Skyrmions we constructed
the corresponding non-rotating compact-star models
in hydrostatic equilibrium; these are mostly
 fluid stars (the Skyrmion fluid) thus naming them  {\it Skyrmion Stars}.
  Here we generalize our previous calculations by
 constructing equilibrium sequences of rotating Skyrmion stars  
  in general relativity using
the computer code ${\it RNS}$ developed by Stergioulas.
We calculated their masses  and radii to be $0.4 \le M/M_{\odot} \le 3.45$,
and   $13.0\ {\rm km}\le R\le 23.0\ {\rm km}$, respectively ($R$ being
 the circumferential radius of the star). The period of the maximally rotating
 Skyrmion stars is calculated to be $0.8\ {\rm ms}\le P \le 2.0\ {\rm ms}$.
We find that a gap (the height between the star surface
and the inner stable circular orbit) starts to appear for $M\sim
2.0M_{\odot}$. Specifically, the Skyrmion star mass range with an existing gap
 is calculated to be  $1.8 < M/ M_{\odot} < 3.0 $
with the corresponding orbital frequency $0.8\ {\rm kHz} < \nu_{\rm ISCO}
< 1.3\ {\rm kHz}$.  We apply our model
to the 4U 1820-30 low mass X-ray binary 
and suggest a plausible 
Skyrmion star candidate in the 4U 1636-53 system. We 
discuss the difficulties  encountered by our model in the 4U
0614+09 case with the highest known Quasi-Periodic Oscillation frequency
of 1329 Hz. A comparative study of Skyrmion stars and  models of neutron
stars based on recent/modern equations of state is also presented.
}

\maketitle

\section{Introduction}
\label{one} 
         
A key input in determining the structure of compact objects 
is the equation of state (EOS) of high density matter whose
 determination  remains a formidable theoretical problem.
There is no general agreement still
on the exact composition of dense matter, and on its EOS, especially for densities in
excess several times nuclear matter density
(in a regime where the EOS is poorly constrained by nuclear data and experiments).
In Ouyed \& Butler (1999, hereafter OB), we presented
and discussed the Skyrme model (Skyrme 1962a\&b)
and its plausible link to Quantum Chromo-Dynamics (QCD).  We explained how the
 connections are intriguing enough that the
model has been revived in order to study its predictions of hadronic interactions.
 Following an approach outlined in K\"albermann (1997), we
thus constructed an equation of state of dense matter which describes a
fluid of Skyrmions coupled to a dilaton field (associated with
the glueball of QCD) and a vector meson field (coupled to the baryon number).
Using our code, we constructed non-rotating,
zero-temperature  stable compact objects --
we called {\it Skyrmion Stars} -- and 
calculated their masses and 
radii to be $0.4 \le M/M_{\odot} \le 2.95$ and 
$11.0\ {\rm km}\le R\le 15.3\ {\rm km}$, respectively.
In \S 5.1 in OB we compared Skyrmion stars -- as defined in our
model --
to other ``Exotic" stars  which also follow from
solutions to an effective non-linear field theory of strong
forces (see also Heusler, Droz, \& Straumann 1992 and
references therein). Skyrmion stars in our picture are not boson/soliton stars
where the soliton is a global structure over the scale of the star but rather
form their constituent baryons as topological solitons
using pions fields. 

In this paper, we  compute models
of rotating Skyrmion stars using the {\it RNS} code written
and made publicly available by N. Stergioulas (Stergioulas \& Friedman 1995).
 {\it RNS} constructs models of {\it rapidly rotating},
{\it relativistic}, {\it compact stars} and assumes {\it uniform rotation}.
The computation solves for the hydrostatic and Einstein
field equations  for uniformly rotating mass distributions,
under the assumptions of stationarity, axial symmetry about the
rotation axis, and reflection symmetry about the equatorial plane. 
The paper is presented as follows: 
In \S \ref{two}, we  describe the equations solved by the {\it RNS} code and
then construct the corresponding rotating Skyrmion stars. 
The results are analyzed and  discussed in \S \ref{three}. A comparative
study of Skyrmion stars and neutron stars is done in \S \ref{four} before 
concluding in \S \ref{five}.

\section{Rapidly and rigidly rotating relativistic Skyrmion stars}
\label{two}

Here we present a brief outline of the equations solved for by
the {\it RNS} code. We start by describing the
space-time around a rotating compact star  in quasi-isotropic coordinates, 
as a generalization of Bardeen's metric (Bardeen 1970):
\begin{eqnarray}
ds^2  =  &-&e^{\gamma +\rho}dt^2 + e^{2\alpha}(r^2d\theta^2+dr^2)\\\nonumber
&+& e^{\gamma-\rho}r^2\sin(\theta)^2 (d\phi -\omega dt)^2\ .
\label{eq1}
\end{eqnarray}
($c=1=G$). The metric potentials,
$\gamma$, $\rho$, $\alpha$, and the angular velocity of the
stellar fluid to the relative local inertial frame ($\omega$)
are all functions of the quasi-isotropic radial coordinates ($r$)
and the polar angle ($\theta$). The function
$(\gamma + \rho)/2$ is the relativistic generalization
of the Newtonian gravitational potential;
the time dilation factor between an observer moving
with angular velocity $\omega$  and an observer
at infinity is $\exp(1/2(\gamma + \rho))$. 
We investigate uniformly rotating perfect fluid
stars with the energy  momentum tensor given by:
\begin{equation}
T^{\mu\nu} = (\epsilon + P)u^{\mu}u^{\nu} + Pg^{\mu\nu}\ ,
\label{eq2}
\end{equation}
where $\epsilon$ is the total energy density, $P$ the pressure
and $u^{\mu}$ the unit time-like four velocity vector that satisfies
\begin{equation}
u^{\mu}u_{\mu} = -1\ .
\label{eq3}
\end{equation}
The proper velocity $v$ of the matter, relative to the local
Zero Angular Momentum Observer (ZAMO), is given in terms of the
angular velocity $\Omega\equiv u^3/u^0$ of the fluid element 
(measured by a distant observer in an asymptotically flat space-time),
by the following equation:
\begin{equation}
v = (\Omega - \omega)r\sin(\theta)e^{-\rho}\ .
\label{eq4}
\end{equation}
The four velocity ($u^{\mu}$) of the matter can be written as
\begin{equation}
u^{\mu} = { e^{-(\gamma + \rho)/2}\over (1-v^2)^{1/2}} (1,0,0,\Omega)\ .
\label{eq5}
\end{equation}
Substitution of the above equation into Einstein field equations projected on to
the frame of reference of a ZAMO yield three elliptic equations for the metric
potentials $\rho$, $\gamma$ and $\omega$ and two linear ordinary differential
equations  for the metric potential $\alpha$ (Bardeen 1971; Butterworth, \&
Ipser 1976; Komatsu, Erigushi, \& Hachisu 1989). In general, the
elliptic differential equations are converted to integral equations for the
metric potentials using Green's function approach.

From the relativistic equations of motion, the equation of
hydrostatic equilibrium for a barytropic fluid (the field
equations are integrated by assuming various zero-temperature, barotropic
EOS of the form $\epsilon = \epsilon (p)$) may be obtained
as :
\begin{eqnarray}
h(P) - h_{\rm p} = {1\over 2} [\gamma_{\rm  p} + \rho_{\rm  p} - \gamma -
\rho &-& \ln(1-v^2)\nonumber \\ &+& A^2 (\omega - \Omega_{\rm  c})^2]\ ,
\label{eq6}
\end{eqnarray}
where $h(P)$ is termed as the specific enthalpy.
$P_{\rm p}$ is  the rescaled values for pressure,
 and $h_{\rm p}$ is the specific enthalpy
at the pole. $\gamma_{\rm  p}$ and $\rho_{\rm  p}$ are the values of
the metric potentials at the pole, and $\Omega = r_{\rm  e}\Omega$.
$A$ is a rotation constant (Komatsu, Erigushi, \& Hachisu 1989) and the
subscripts, $p$, $e$ and $c$ stand for polar, equatorial and central,
respectively. {\it RNS} solves the integral equation for $\rho$, $\gamma$ and
$\omega$, the ordinary differential equation (in $\theta$) for the metric
potential $\alpha$, together with Eq.~(\ref{eq6}), the hydrostatic equilibrium
equations at the center [$h(P_{\rm c})$ given] and at the equator [$h(P_{\rm
e})= 0$], iteratively to obtain $\rho$, $\gamma$, $\alpha$, $\omega$, the
equatorial coordinate radius ($r_{\rm  e}$), angular velocity ($\Omega$), and
the density ($\epsilon$) and pressure($P$) profiles.

The geodesic motion for circular orbits around
rotating relativistic stars is also presented in Bardeen (1970)
 and  correspond to the motion of ZAMO yielding two possible values
 for the velocity corresponding to corotating and counter-rotating
orbits. The innermost stable circular orbits are
given by solving for $V,rr = 0$ where $V$ is the effective potential as
written in Bardeen (1972);  the comma followed by $rr$
represent a second order partial derivative with respect to radius $r$. 

\section{Numerical results}
\label{three}

\subsection{Maximally rotating configurations}

The highest 
energy density used in all the computed rotating
models is the value which gives the
maximum mass non-rotating star as calculated in OB.
We specify  the equation of state,
and the central energy density and the code computes models with
increasing angular velocity until the star is spinning with 
the same angular velocity as a particle orbiting the star
at its equator\footnote{We will
sometimes refer to it as the Kepler frequency which is the frequency of
 a particle in a stable orbit at the circumference
 of a star. It is therefore also the frequency at which
 disruption of the star would occur.}. 

\begin{figure}[t!]
\centerline{\includegraphics[width=0.5\textwidth]{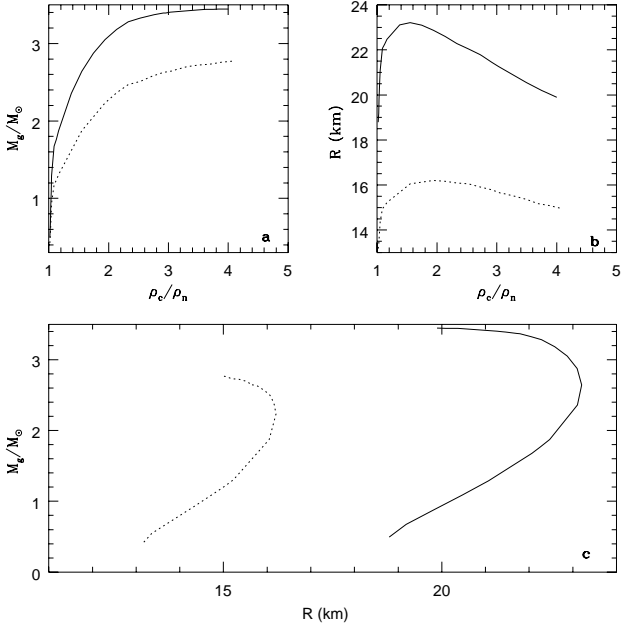}}
\caption{
{\bf Maximally rotating configurations}: {\bf a).}
Gravitational  Mass versus central density (in units of the
nuclear saturation density) for zero
temperature Skyrmion stars in hydrostatic equilibrium. In this figure, the
solid curves represent the rotating configurations while the dotted curves
represent the non-rotating or static configurations. {\bf b).} 
Radius versus central density. {\bf c).} The Mass$-$Radius plane.
}
\label{Fig.1}
\end{figure}

\begin{figure}[t!]
\centerline{\includegraphics[width=0.5\textwidth]{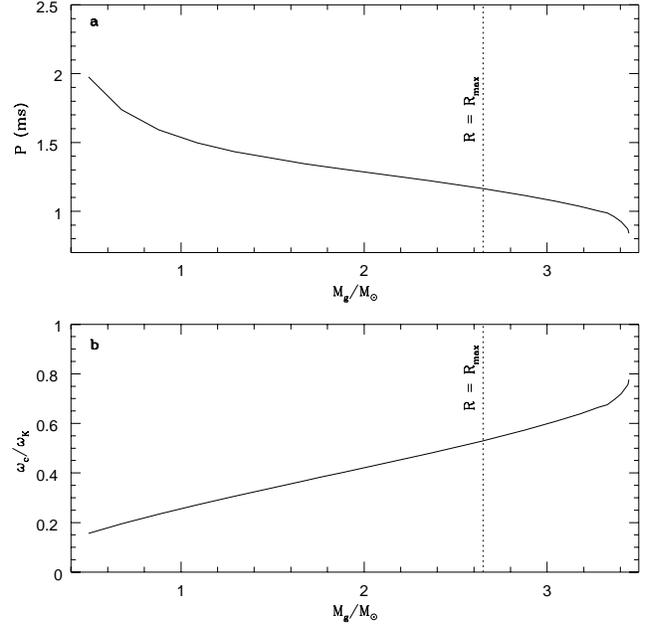}}
\caption{
{\bf Maximally rotating configurations}: {\bf a).} Period
versus  gravitational mass for zero temperature
Skyrmion stars in hydrostatic equilibrium. {\bf b).} Ratio of the inertial frame
dragging at the center of the star to the rotation rate versus gravitational
mass. In this Figure and the subsequent ones, the  vertical dotted line
illustrates  the  mass of the maximum radius configuration.
}
\label{Fig.2}
\end{figure}

Fig.~\ref{Fig.1}a and 1b show the resulting stellar masses and radii as a
function of the central density, respectively.  The mass range is
$0.4\le M/M_{\odot}\le 3.45$ while the corresponding radii (equatorial
circumferential radius; [proper equatorial circumference]/2$\pi$) are
calculated to be  $18.6\ {\rm km}\leq R\leq 23.0\ {\rm km}$. 
In Fig.~\ref{Fig.1}c, we show the resulting Mass-Radius plane. 
In general, for a given central density, rotation allows
the masses and radii to increase by  30 \% and 40\%, respectively,
when compared with the non-rotating cases. 
The amount of baryonic mass in the outer region of the star
(the crust region where $\rho < \rho_{\rm  N}$ is constructed
 using the EOS of Baym, Pethick, \& Sutherland (1971); $\rho_{N}$
is the nuclear saturation density)
decreases drastically with rotation.  Rotating Skyrmion stars crust
constitute less than 5\% of the total baryonic mass  while
it averages 20\% for the static configurations. The   nuclear crust
is further addressed in OB to which we refer the interested reader.

The minimum spin period as a function of mass is  
shown in Fig.~\ref{Fig.2}a and is calculated to be 
  $0.8\ {\rm ms}\le P \le 2.0\ {\rm ms}$.
The  inertial frame dragging at the center of the star to the rotation rate is
plotted in Fig.~\ref{Fig.2}b which shows a significant value throughout and
reaches a maximum value of $0.8$ for the most massive stars. It suggests
that frame-dragging might have a significant effect
on the structure of rotating Skyrmion stars as $P$ decreases
since it is taking place against the background of a radially
dependent, frame-dragging frequency (see Weber \& Glendenning 1992 for
example).

It has been noted by H{\ae}nsel \& Zdunik (1989) and Friedman \& Ipser (1992)
that  the maximum spin rate for many  neutron star EOS seems to be given by
\begin{equation}
\Omega_{max} = \chi ({M_{\rm s, max.}\over M_{\odot}})^{1/2}
({R_{\rm s, max.}\over 10\ {\rm km}})^{-3/2}\quad s^{-1}\ ,
\label{eq7}
\end{equation}
where $\chi$ has been quoted as either 7600 or 7700 $s^{-1}$.
$M_{\rm s, max.}$ and $R_{\rm s, max.}$ are the total mass-energy and areal
radius of the maximum-mass static configuration for a given EOS. The newest 
fit for the proportionality constant can be found in  Haensel, Salgado,
\& Bonazzola (1995) 
where $\chi = 7600{\rm -}7900\ {\rm s^{-1}}$
is given.  If we assume that the minimum spin period for the Skyrmion star is
0.8 ms (that is the star with the maximum mass on the Kepler sequence
has the maximum rotation rate) then the maximum angular velocity is 7854 rad
$s^{-1}$. Using  the values for the maximum mass spherical star we get,
\begin{eqnarray}
\Omega_{max} &=& 7600{\rm -}7900\times (2.95)^{1/2} (1.44)^{-3/2}\nonumber\\
&=& 7555{\rm -}7852~s^{-1}\ .
\label{eq8}
\end{eqnarray}
This corresponds to less than few percent error which makes
Eq.~(\ref{eq7})  an excellent tool for  predicting the maximum angular
velocity even for our EOS.

\begin{figure}[h!]
\centerline{\includegraphics[width=0.5\textwidth]{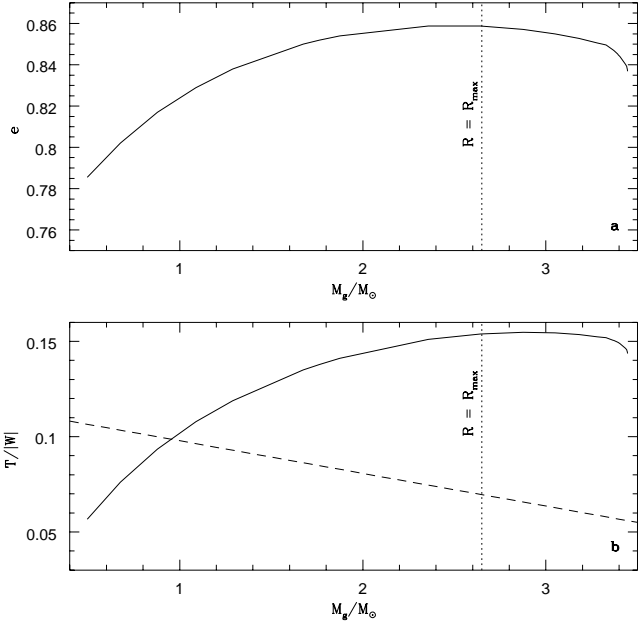}}
\caption{
{\bf Maximally rotating configurations}: {\bf a).}
Eccentricity   versus gravitational mass for zero temperature
Skyrmion stars in hydrostatic equilibrium. {\bf b).}  Kinetic to gravitational
 energy ratio versus gravitational mass. The dashed line 
is based on  the empirical formula for the onset of the
bar mode instability as described by Eq.\ref{eq9}.
Configurations with $T/\vert W\vert $  above the dashed line  might be
unstable to the bar mode.}
\label{Fig.3}
\end{figure}

The centrifugal flattening of maximally rotating
Skyrmion stars is obvious
from Fig.~\ref{Fig.3}a with eccentricities $e=\sqrt{1-(R_p/R_e)^2}$
 ($R_p$ and $R_{\rm e}$ being the polar
and equatorial radius of the configuration, respectively) ranging
from $0.78$ for the  lightest stars up to $0.86$
for the heaviest ones. The maximum eccentricity is reached at $M= 2.62 M_{\odot}$
(the maximum radius configuration) after which $e$ decreases,  as expected.
Fig.~\ref{Fig.3}b shows the Kinetic to gravitational
 energy ratio ($T/\vert W\vert$) versus gravitational mass.
 Configurations with $T/\vert W\vert $ 
above the dashed line  might be unstable to the bar mode. 
We come back to this  in \S 3.2 where we
 discuss the conditions for the onset of 
non-axisymmetric modes and to what extent they
 could be  used as a stronger constraint
than the maximum rotational frequency. 

\begin{figure}[t!]
\centerline{\includegraphics[width=0.5\textwidth]{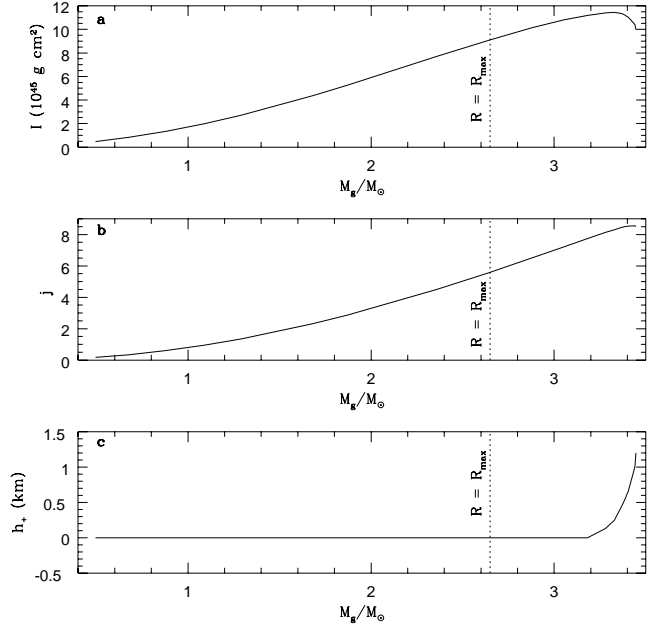}}
\caption{
{\bf Maximally rotating configurations}: {\bf a).} Moment
of inertia  versus gravitational mass for zero temperature
Skyrmion stars in hydrostatic equilibrium. {\bf b).} Angular momentum 
(Dimensionless ratio of angular
momentum $cJ/GM_{\odot}^2$) versus gravitational mass. 
{\bf c).} Height from surface of
 last stable co-rotating orbit
in equatorial plane versus gravitational mass.
}
\label{Fig.4}
\end{figure}

In Fig.~\ref{Fig.4}, the moment of inertia (Fig.~\ref{Fig.4}a) and 
the angular momentum (Fig.~\ref{Fig.4}b) are plotted.
The sudden decrease in the moment of inertia
for $M \ge 3.3 M_{\odot}$ is the consequence of the
 mass reaching a ``plateau" (at $M = 3.3 M_{\odot}$) while
the radius keeps decreasing (see Fig.~\ref{Fig.1}a and Fig.~\ref{Fig.1}b).
The combined effects of $I$ decreasing with $\Omega$ increasing
(Fig.~\ref{Fig.2}a) explains then why
the angular momentum stays constant for
 $M \ge 3.3 M_{\odot}$ - using the simple dimensional analysis $J = I\Omega$. 
Finally the height from surface of the last stable co-rotating orbit
 ($h_{+} = r_{\rm orb.} - R$)\footnote{For the cases where $r_{\rm orb.}$ is
non-existent or inside the star, $r_{\rm orb.}$ is taken to be the
Keplerian orbit radius at the surface of the star.} is shown in
Fig.~\ref{Fig.4}c. It can be seen from this figure 
that in general the stable orbits exists 
 up to the surface of the star. The large radii calculated
  makes it very difficult for
the stable orbit to be outside the star (when compared to $6GM/c^2$; the
 inner most stable Kepler orbit around the non-rotating case).
For the most massive configurations ($M \ge 3.2 M_{\odot}$),
the boundary layer (the separation between the
surface of the neutron star and its innermost stable orbit) can be as
as high as 1.5 km for the maximum value. Lowering the spin frequency will
eventually relax the above conclusions (see \S 3.3).

\subsection{Bar instability}

Skyrmion stars are
subject to rotational constraints
linked to gravitational-wave instabilities  which make all rotating,
 perfect fluid equilibria unstable
to modes with angular dependence $\exp(im\phi)$ for sufficiently
large $m$ (Friedman \& Schutz 1978; Lindblom 1984).
 These  instabilities allow the star to convert
its rotational energy to gravitational energy waves (Glendenning 1997).
 
Fairly recent general relativistic calculations have shown that the 
gravitational-radiation-driven bar-mode ($f$-mode) instability
sets in at values of $T/\vert W\vert $ much lower than in the Newtonian limit.
 For the case of realistic EOS, Morsink, Stergioulas, \&
Blattnig (1999) found the following empirical formula for the
onset of the bar mode instability (the case of polytropes
can be found in Stergioulas \& Friedman 1998): \begin{equation}
T/\vert W\vert  = 0.115 - 0.048 M/M^{sph}_{max}\ ,
\label{eq9}
\end{equation}
where $M^{sph}_{max}$ is the maximum mass star for the 
non-rotating case. Assuming that this formula holds for our
EOS, we find that all maximally rotating stars
with $M\ge M_\odot$ might be prone to such an instability.

The real question, however, is whether the $f$-mode instability is 
actually going to limit the spin period of Skyrmion stars. With 
the onset of the $m=2$ ($f$-mode),  the 
r-mode\footnote{The $r$-mode instability is a member of the class of
gravitational-radiation-driven instabilities and are generic
in rotating neutron stars (Friedman \& Morsink 1998).
That is, every $r$-mode is in principle unstable in 
every rotating star, in the absence of viscosity.}
 will soon become unstable (Friedman \& Morsink 1998) and
spin down the star to even slower rate. As a result, it is not
expected that the $f$-mode will limit 
the spin-rate of any stars (Andersson, Kokkotas, \& Schutz 1999),
 although it may become unstable and emit gravitational
radiation in some cases.

\subsubsection{Skyrmion fluid viscosity}

Realistic compact stars are viscous, and the presence of
viscosity will shift the onset of instability and if large enough
 will damp out the instability (Andersson, Kokkotas, \& Schutz 1999).
If the instability is driven by viscous dissipation,
 it was shown in Bonazzola, Frieben, \& Gourgoulhon (1995) 
 that for cold uniformly rotating NS only very stiff
equations of state are likely to allow for spontaneous symmetry breaking.
If we consider the stiffness of our EOS (when
compared to standard EOS;  see Figure 2 in OB)
then instability to a bar mode is a very likely
plausibility in Skyrmion stars.  So how viscous is a Skyrmion fluid
and what is its temperature dependency?

The Skyrmion fluid can be looked at as made of fermionic soliton objects
and in principle one should be able to calculate
its physical parameters, among others its viscosity.
The complication is linked to the fact that
unlike a pure fermionic fluid, the structure
of the Skyrmion fluid is highly
non-linear (a consequence of the Skyrmion having structure).
 This makes the calculations not trivial
and beyond the scope of this paper.
There might still be  the possibility  
 that such a fluid is viscous enough that it will of course damp these
instabilities altogether so that limits on rotation is set
by the Kepler frequency - a notion to be confirmed.

\begin{figure}[t!]
\centerline{\includegraphics[width=0.5\textwidth]{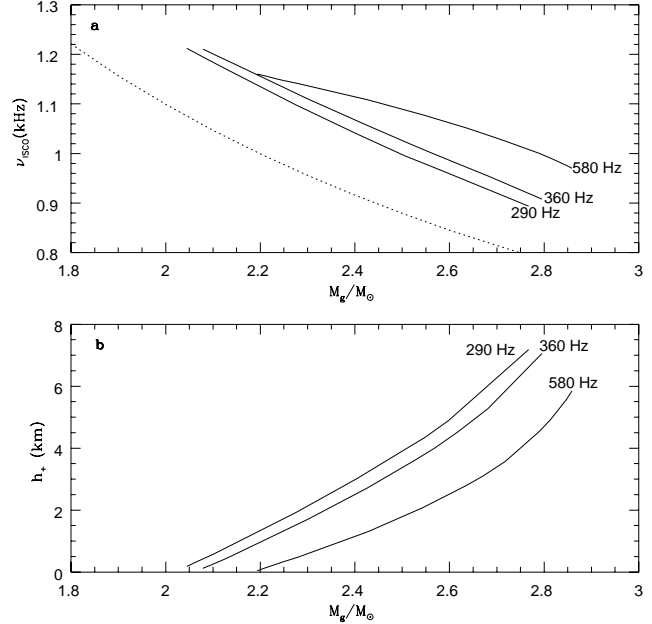}}
\caption{
{\bf Different spin frequencies}: 
{\bf a).} The  ISCO frequency
versus mass for Skyrmion stars rotating at 290, 360 and 580 Hz respectively.
The dotted line represents the static models, $\nu_{ISCO}=
2198\ {\rm Hz} (M_{\odot}/M)$. {\bf b).} 
Height from the star surface of the innermost stable circular orbit 
(ISCO) versus mass for
Skyrmion stars rotating at 290, 360 and 580 Hz respectively (see Table 1).
}
\label{Fig.5}
\end{figure}

\subsection{Different spin frequencies: Skyrmion stars and QPOs}

The discovery of kilohertz Quasi-Periodic
 Oscillations (QPOs) in low mass X-ray binaries
(LMXBs) with the {\it Rossi X-Ray Timing Explorer} has stimulated extensive
studies of these sources (van der Klis 1998). 
One thing that has been suggested in the literature is that the highest frequency
QPO observed could correspond to the orbital frequency at the
inner stable circular orbit (ISCO) (Miller, Lamb, \& Psaltis 1998).
The signature of the marginally stable  orbit
is a saturation in QPO frequency (assumed to
track inner disk radius) versus mass accretion.
Such a saturation where the frequency becomes
independent of the chosen mass accretion rate
indicator  has been reported in 4U 1820-30
(Zhang et al. 1998; Kaaret et al. 1999).

\begin{table*}[!ht]
\caption{Skyrmion stars: Different spin frequencies ($\nu_{s}$)}
\begin{center}
\begin{tabular}{ccccccccccc}
\hline
$\epsilon_{c}$ & $M_{G}$ & $M_{0}$ & $R$ & $h_{+}$ & $h_{-}$ & $e$ & $I$ & $j$
& $T/W$ & $\omega_{c}/\Omega_{c}$\\ \hline
\multicolumn{11}{c}{$\nu_{s}$ = 290.0 Hz}\\
\hline
3.0$\times 10^{14}$ & 1.35 & 1.46 & 15.67 & 0.00 & 0.00 & 0.31 & 2.36 & 0.49 & 0.016 & 0.31 \\
5.0$\times 10^{14}$ & 2.28 & 2.59 & 16.47 & 1.93 & 6.26 & 0.28 & 5.09 & 1.06 & 0.011 & 0.49 \\
6.9$\times 10^{14}$ & 2.60 & 3.02 & 16.13 & 4.89 & 9.31 & 0.24 & 5.83 & 1.21 & 0.009 & 0.59 \\
\hline
\multicolumn{11}{c}{$\nu_{s}$ = 360.0 Hz}\\
\hline
3.0$\times 10^{14}$ & 1.38 & 1.48 & 15.93 & 0.00 & 0.00 & 0.39 & 2.48 & 0.64 & 0.025 & 0.31 \\
5.0$\times 10^{14}$ & 2.30 & 2.62 & 16.62 & 1.69 & 7.08 & 0.34 & 5.24 & 1.35 & 0.017 & 0.50 \\
6.9$\times 10^{14}$ & 2.62 & 3.04 & 16.24 & 4.53 & 10.04 & 0.28 & 5.96 & 1.53 & 0.014 & 0.59\\
1.0$\times 10^{15}$ & 2.80 & 3.29 & 15.26 & 7.06 & 12.18 & 0.24 & 5.81 & 1.50 & 0.010 & 0.69\\
\hline
\multicolumn{11}{c}{$\nu_{s}$ = 580.0 Hz}\\
\hline
3.0$\times 10^{14}$ & 1.53 & 1.66 & 17.43 & 0.00 & 3.33 & 0.64 & 3.25 & 1.35 & 0.069 & 0.34 \\
5.0$\times 10^{14}$ & 2.43 & 2.76 & 17.42 & 1.31 & 10.17 & 0.53 & 6.05 & 2.51 &0.048 & 0.52 \\
6.9$\times 10^{14}$ & 2.72 & 3.16 & 16.78 & 3.58 & 12.72 & 0.47 & 6.58 & 2.73 & 0.038 & 0.61\\
1.0$\times 10^{15}$ & 2.86 & 3.36 & 15.57 & 5.85 & 14.29 & 0.39 & 6.18 & 2.56 & 0.028 & 0.70\\
\hline
\end{tabular}
\end{center}
\end{table*}
\normalsize

  In Fig.~\ref{Fig.5}a we plotted ISCO frequency (for the
prograde case only)  versus Mass for 
 Skyrmion star models rotating at frequencies 290, 360 and 580 
Hz respectively.
These values allow us to cover the range of frequencies observed in LMXBs.
Fig.~\ref{Fig.5}b shows the corresponding gaps; that is the height between the
star surface and the ISCO (prograde case).   In general,  the Skyrmion star
mass range with an existing gap  is calculated to be,  
\begin{equation}
1.8\ {\rm M_{\odot}} < M < 3.0\ {\rm M_{\odot}}\ ,
\label{eq10}
\end{equation}
while the corresponding radii and $\nu_{ISCO}$ are
\begin{equation}
14.0\ {\rm km} < R < 20.0\  {\rm km}\ ,
\label{eq11}
\end{equation}
\begin{equation}
0.8\ {\rm kHz} < \nu_{\rm ISCO} < 1.3\ {\rm kHz}\ ,
\label{eq12}
\end{equation}
respectively. 
Our numbers can thus account for 
the high masses (up to $1.78\pm 0.23M_{\odot}$; Orosz, \& Kuulkers 1999)  
 determined from LMXBs (see also Miller, Lamb, \& Cook 1998). 
General features of the  models discussed in this
section are shown in Table 1. A key to all the the tables in
this paper is as follows:\\

\begin{tabular}{ll}
$\Omega_{s}$&Star's angular velocity ($10^4$ Hz)\\
$\nu_{s}$&Star's spin frequency, $\nu = \Omega/2\pi$ (Hz)\\
$\epsilon_{c}$&Central density (g cm$^{-3}$)\\
$M_{G}$&Gravitational mass ($M_{G}/M_{\odot}$)\\
$M_{0}$&Baryonic mass ($M_{0}/M_{\odot}$)\\
$R$ &Radius of star, measured at equator (km)\\
$r_{orb}$&Radius  of the innermost stable orbit (km)\\
$h_{+}$&Height from the star surface of innermost 
stable\\
 &prograde circular orbit (km)\\
$h_{-}$&Height from the star surface 
of innermost stable\\
 &retrograde circular orbit (km)\\ 
$e$&Eccentricity\\
$I$&Moment of Inertia ($10^{45}$ g cm$^{2}$)\\
$j$&Angular momentum  ($cJ/GM_{\odot}^2$)\\
$T/\vert W\vert $&Rotational energy to the gravitational energy\\
$\omega_{c}/\Omega_{c}$&Ratio of the inertial frame dragging
 at the\\
 &center of the star to the rotation rate\\
$Z_{p}$&Polar redshift\\
$Z_{f}$&Forward redshift\\
$Z_{b}$&Backward redshift
\end{tabular}

To first order in the
rotation rate of the star ($\nu_{s} < 400$ Hz), the orbital frequency in the
marginally stable orbit is $\nu_{ISCO} \simeq 2198\ {\rm Hz}
(M_{\odot}/M)(1+0.748j(M_{\odot}/M)^2)$ (see Miller, Lamb, \& Cook 1998,
for example).
Consider, as an example, a moderately slowly rotating
Skyrmion star, $\nu = 290$ Hz, with mass $M = 2.6M_{\odot}$. Using the value
of $j$ given in Table 1 we find $\nu_{ISCO}\sim 9589$ Hz
while the value in the non-rotating case is 
$8454$ Hz (depicted by the dotted line in Fig.~\ref{Fig.5}a).

\subsubsection{The case of {\bf 4U 1820-30}}

We first apply our model to 
4U 1820-30 which seems to present the strongest experimental evidence
for the existence of the marginally stable orbit 
(Zhang et al. 1998; Kaaret et al. 1999).
For $\nu_{\rm ISCO} \simeq 1060$ Hz as the estimated frequency
at the ISCO, and the spin frequency to near
300 Hz, Fig.~\ref{Fig.5}a suggests a mass
 2.35$M_{\odot}$.    
The radius is then calculated to be of the order of 16 km.
These numbers remain to be confirmed from observations and furthet
modelling of the system. 

\subsubsection{The case of {\bf 4U 0614+09}}

The highest known QPO has a frequency
of 1329 Hz (van Straaten et al. 2000).
The frequency of the corresponding star 
(4U 0614+09) is believed to be near 300 Hz.
Simple estimates 
(van Straaten et al. 2000, \S 4.4; van der Klis 2000, \S 5.6)
show the difficulty for extremely stiff EOS  such
as ours to account for the derived range in mass and radius
($M < 1.9 M_{\odot}$ and $R < 15.2$ km). 
We have noted that for our EOS $\nu_{ISCO} < 1.3$ kHz and
$R >$ 14 km. Recent modeling
put even more constraints by suggesting a  
 more compact 4U 0614+09 with a radius less than
10 km (Titarchuk \& Osherovich 2000).

It is not trivial to account for such numbers in our model 
unless  the 1329 Hz frequency does not actually correspond to the
ISCO. 
Indeed, unlike in the 4U 1820-30 source,
there is no obvious
saturation of the kilo-hertz QPO with respect to 
mass accretion indicator in the 4U 0614+09
case (see Fig.4 in van Straaten et al. 2000). 

\section{A comparative study of Skyrmion stars and Neutron stars}
\label{four}

In this chapter we compare Skyrmion stars 
to the more traditional neutron stars. We chose three recent/modern EOSs:
the first is described  in Baldo, Bombaci, \& Burgio (1997)
which   uses the Argonne v14 (Av14)) (Wiringa, Smith, \& Ainsworth 1984)
 two-body nuclear force; the second uses the Paris
two-body nuclear force (Lacombe et al. 1980) , implemented in
both cases by the Urbana three-body force  (TBF)
(Carlson, Pandharipande, \& Wiringa 1983; Schiavilla, Pandharipande, \&
Wiringa 1986) while the third one is that
developed   in Prakash et al. (1997)
 using a generalized Skyrme-like EOS. As in
Datta, Thampan, \& Bombaci (1998), we refer to these EOSs as 
BBB1 (Av14+TBF), BBB2 (Paris+TBF) and BPAL32, respectively.

\begin{figure}[t!]
\centerline{\includegraphics[width=0.5\textwidth]{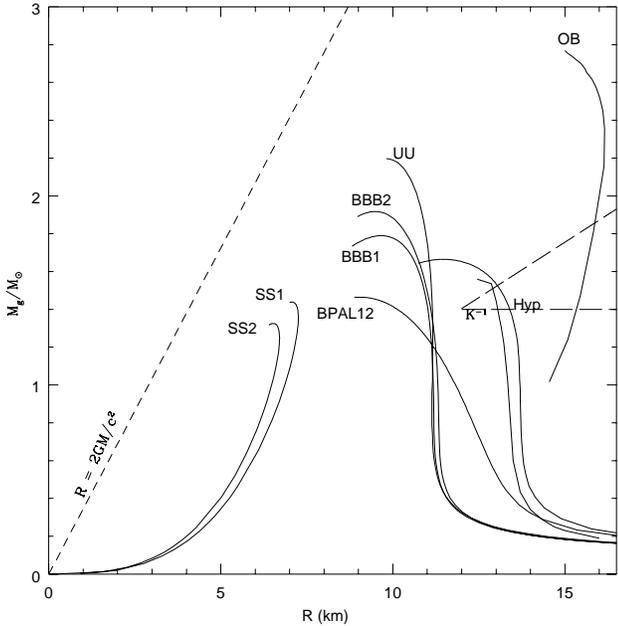}}
\caption{
The $M-R$ relation for non-rotating Skyrmion
stars (OB) as compared to  theoretical models of non-rotating neutron stars 
(UU, BBB1, BBB2, BPAL12, Hyp, and K$^{-1}$) and
strange stars (SS1 and SS2). The data for the  
neutron stars and strange stars was kindly provided to us by Li et al. (1999). 
The  Schwarzschild radius ($2GM/c^2$) is shown as a dotted line. 
Inside the triangle is the allowed range of $M$ and $R$ for 4U 1636-53 as
modeled in Nath, Strohmayer, \& Swank (2001) using fits to X-ray bursts.}
\label{Fig.6}
\end{figure}

\begin{table}[!h]
\caption{Maximum-mass non-rotating models}
\begin{tabular}{cccccc}
\hline
EOS & $\epsilon_{c}$ & $M_{G}$ & $M_{0}$ & $R$ &  $r_{orb}$\\
\hline
BBB1   & 3.09$\times 10^{15}$ & 1.788 & 2.082 & 9.646 &  15.845\\
BBB2   & 3.12$\times 10^{15}$ & 1.917 & 2.261 & 9.519 & 16.984\\
BPAL32 & 2.67$\times 10^{15}$ & 1.947 & 2.263 & 10.509& 17.254\\
OB     & 1.22$\times 10^{15}$ & 2.797 & 3.299 & 14.565& 24.743\\
\hline
\end{tabular}
\end{table}

\begin{table*}[!ht]
\caption{Maximum-mass rotating models}
\begin{center}
\begin{tabular}{cccccccccc}
\hline
EOS & $\epsilon_{c}$ & $M_{G}$ & $M_{0}$ & $R$ & $r_{orb}$ & $e$ &
$\Omega_{s}$\\ \hline
BBB1   & 2.56$\times 10^{15}$ & 2.135 & 2.471 & 13.129 & 13.490 & 0.703 & 1.095\\
BBB2   & 2.82$\times 10^{15}$ & 2.272 & 2.653 & 12.519 & 13.550 & 0.687 & 1.203\\
BPAL32 & 2.27$\times 10^{15}$ & 2.300 & 2.657 & 14.276 & 14.611 & 0.699 & 1.001\\
OB     & 1.03$\times 10^{15}$ & 3.445 & 4.034 & 19.890 & 21.089 & 0.837 & 0.749\\
\hline
 ...  & $I$ & $j$ & $T/W$ & $\omega_{c}/\Omega_{c}$ &  $Z_{p}$ & $Z_{f}$ & $Z_{b}$\\
\hline
...   & 2.428 & 3.019 & 0.120 & 0.764  & 0.690 & -0.330 & 1.975\\
...   & 2.539 & 3.469 & 0.123 & 0.825  & 0.849 & -0.349 & 2.483\\
...   & 3.005 & 3.416 & 0.113 & 0.771  & 0.679 & -0.328 & 1.933 \\
...   & 9.982 & 8.507 & 0.144 & 0.777  & 0.777 & -0.342 & 2.287\\
\hline
\hline
\end{tabular}
\end{center}
\end{table*}

\begin{table*}[!ht]
\caption{Maximum angular momentum  models}
\begin{center}
\begin{tabular}{cccccccccc}
\hline
EOS & $\epsilon_{c}$ & $M_{G}$ & $M_{0}$ & $R$ & $r_{orb}$ & $e$ &
$\Omega_{s}$\\ \hline
BBB1   & 2.44$\times 10^{15}$ & 2.133 & 2.468 & 13.264 & 13.558 & 0.706 & 1.079\\
BBB2   & 2.82$\times 10^{15}$ & 2.272 & 2.653 & 12.519 & 13.550 & 0.687 & 1.203\\
BPAL32 & 2.14$\times 10^{15}$ & 2.299 & 2.655 & 14.481 & 14.713 & 0.702 & 0.981\\
OB     & 9.00$\times 10^{14}$ & 3.435 & 4.017 & 20.549 & 21.498 & 0.840 & 0.715\\
\hline
... & $I$ & $j$ & $T/W$ & $\omega_{c}/\Omega_{c}$ &  $Z_{p}$ & $Z_{f}$ & $Z_{b}$\\
\hline
...    & 2.465 & 3.021 & 0.120 & 0.756 & 0.677 & -0.328 & 1.935\\
...    & 2.539 & 3.469 & 0.123 & 0.825 & 0.849 & -0.349 & 2.483\\
...    & 3.070 & 3.419 & 0.114 & 0.760 & 0.661 & -0.326 & 1.881\\
...    & 10.52 & 8.560 & 0.146 & 0.751 & 0.728 & -0.355 & 2.133\\
\hline
\end{tabular}
\end{center}
\end{table*}
\normalsize
Table 2 summarizes the non-rotating compact star structure parameters
for the EOS models BBB1, BBB2, BPAL32 and ours (OB). 
The values listed correspond to the maximum stable mass configuration.
Note that for this table and the rest of the tables, we consider
only the co-rotating case (the cases where $r_{\rm orb.}$ is
non-existent or inside the star, $r_{\rm orb.}$ is taken to be the
Keplerian orbit radius at the surface of the star). 
The maximum mass is usually an indicator of
the softness/stiffness of the EOS and its  values as listed in Table 2
confirms our previous statements  that our EOS is
stiffer than the standard ones. The consequences
of such a stiffness is also illustrated in 
Table 3 where we list the  quantities corresponding to 
the maximum gravitational mass configurations.  
It can be seen that the gravitational mass
of the maximum stable rotating Skyrmion star has a value of $\sim 3.5M_{\odot}$,
while it is $\sim 2.4M_{\odot}$ for the configurations
constructed with the other EOS. That is, with our EOS we do not need to go to
break up speeds to account for the $\sim 2.0M_{\odot}$ mass  predicted from
analysis of LMXB observational data (Zhang et al. 1996).
The values of these same  quantities are listed in Table 4 for the maximum
angular momentum models leading to similar conclusions.

\subsection{{\bf 4U 1636-53} as a Skyrmion star candidate?}

In Fig.~\ref{Fig.6} we compare the $M-R$ relation for Skyrmion
stars (OB) to the  theoretical $M-R$ curve obtained using six recent
realistic models for the EOS (UU, BBB1, BBB2, BPAL12, Hyp, and K$^{-1}$).
The solid curves labeled SS1 and SS2 are for 
strange stars (the data was kindly provided to us by Li et al. 1999). 
The triangle depicts the mass-radius constraint from fits to X-ray bursts
in 4U 1636-53. Inside the triangle is the
allowed range of $M$ and $R$ which satisfies the compactness constraints
as modeled in Nath, Strohmayer, \& Swank (2001; see their Figure 4), and
clearly favoring stiffer EOSs. Our modeled stars (OB) cross the triangle
suggestive of 4U 1636-53 as a plausible Skyrmion star candidate.
However, one should keep in mind the fact that modern EOSs can be modified
as to also cross the triangle (Heiselberg \& Hjorth-Jensen 1999).

\section{Conclusion}
\label{five}

With the {\it RNS} code, we constructed numerical models of rotating
Skyrmion stars for a newly derived  EOS of dense matter based on a fluid of Skyrmions. 
We calculated their masses and radii to be $0.4 \le M/M_{\odot} \le 3.45$ and    
 $13.0\ {\rm km}\le R\le 23.0\ {\rm km}$, respectively.
 The spin period of the maximally rotating configurations are calculated to be  
  $0.8\ {\rm ms}\le P \le 2.0\ {\rm ms}$.
Owing to the stiffness of the Skyrmion fluid,  Skyrmions stars are 
found to be on average heavier, to show higher equatorial radii,
and to rotate slower than theoretical models of neutron stars based on modern
realistic EOS.

Skyrmion stars might have little bearing on reality
since there is still
no guarantee that our EOS (OB)
is a plausible representation of matter above nuclear densities. However,
our model so far has been successful in reproducing the basic features of
compact objects.  
An interesting consequence of our model, with its plausible
applications to QPO systems, is the fact that massive Skyrmion stars
($1.8 < M/ M_{\odot} < 3.0 $) can possess gaps with
orbital frequencies in the kHz range ($0.8\ {\rm kHz} < \nu_{\rm ISCO}
< 1.3\ {\rm kHz}$). These points
suggest that the model warrants further study.

\begin{acknowledgements}
I am grateful to  S. Morsink, M. Butler 
and G. K\"albermann for encouraging help and valuable
discussions. I am also grateful to an anonymous referee for the remarks that
helped improve this work. My thanks to  N. Stergioulas for making his code
available. 

\end{acknowledgements}

\end{document}